\documentclass{ws-ijmpcs}

\begin{document}

\markboth{B. Liu (for BESIII collaboration)}
{Partial Wave Analysis at BESIII}

\catchline{}{}{}{}{}
%
%%%%%%%%%%%%%%%%%%%%%%%%%%%%%%%%%%%%%%%%%%%%%%%%%%%%%%%%%%%%%%%%%%%%

\title{Partial Wave Analysis at BESIII}

\author{Beijiang Liu (for BESIII collaboration)}

\address{Institute of High Energy Physics, Chinese Academy of Sciences, 19B Yuanquanlu, Shijingshan district\\
Beijing, 100049,
China\\
liubj@ihep.ac.cn}

\maketitle

\begin{history}
\received{Day Month Year}
\revised{Day Month Year}
\end{history}

\begin{abstract}
The BESIII experiment in Beijing takes data in $\tau$-charm domain since 2009. For the moment the world largest samples of $J/\psi$, $\psi(3686)$, $\psi(3770)$ and $\psi(4040)$ data have been collected. Hadron spectroscopy is a unique way to access QCD, which is one of the most important physics goals of BESIII. Experimental search of new forms of hadrons and subsequent investigation of their properties would provide validation of and valuable input to the quantitative understanding of QCD. The key to success lies in high levels of precision during the measurement and high statistics in the recorded data set complemented with sophisticated analysis methods. Partial wave analysis (PWA) is a powerful tool to study the hadron spectroscopy, that allows one to extract the resonance's spin-parity, mass, width and decay properties with high sensitivity and accuracy.
In this poster, we present the working PWA framework of BESIII -- GPUPWA and the recent results of PWA of $J/\psi\to\gamma\eta\eta$. GPUPWA is a PWA framework for high statistics partial wave analyses harnessing the GPU parallel computing.

\end{abstract}

\ccode{PACS numbers: 13.25.Gv, 13.40.Hq, 14.40.Be}

\section{Introcduction}	
Hadron spectroscopy is a unique way to
access QCD. This theory predicts that
there should exist new forms of matter,
such as glueballs (pure-gluon objects)
and hybrids ($q\bar{q}$ states with explicit
gluon)
\cite{Close:1987er}\cdash\cite{Crede:2008vw}. Experimental search of these
predictions and subsequent
investigation of their properties would
provide validation of and valuable input
to the quantitative understanding of
QCD. From the experimental side, the
basic task is to systematically map out
all the resonances with the
determination of their properties like
mass, width, spin-parity as well as
partial decays widths.

BESIII (Beijing Spectrometer)is a new
state-of-the-art $4\pi$ detector at the
upgraded BEPCII (Beijing Electron and
Positron Collider) that operated in the $\tau$-charm threshold energy region\cite{bes3dect}.
Since 2009, it has collected the world¡¯s
largest data samples of $J/\psi$, $\psi(3686)$,
$\psi(3770)$ and $\psi(4040)$ decays. These data
are being used to make a variety of
interesting and unique studies of light
hadron spectroscopy precision
charmonium physics and high-statistics
measurements of D meson decays\cite{Asner:2008nq}.

The study of radiative decays of $J/\psi$ is
considered most suggestive in the
glueball search. After photon emission,
the $c\bar{c}$ annihilation can go through $C-even$
$gg$ states, and hence may have a
strong coupling to the low-lying
glueballs \cite{Close:1987er}\cdash\cite{Crede:2008vw}.

\section{Partial wave analysis method}

Extracting resonance properties from experimental data is however far from
straightforward; resonances tend to be broad and plentiful, leading to intricate
interference patterns. In such an environment, simple fitting of mass spectra is
usually not sufficient and a partial wave analysis (PWA) is required to disentangle
interference effects and to extract resonance properties. In the cases discussed here,
the full kinematic information is used and fitted to a model of the amplitude in a
partial wave decomposition. The partial wave amplitude is constructed with an
angular part and a dynamical part. The model parameters are determined by an
unbinned likelihood fit to the data, while the event-wise efficiency correction is
included. In a typical PWA (we use the radiative decay $J/\psi\to\gamma\eta\eta$\cite{Ablikim:2013hq} as an example),
the quasi two-body decay amplitudes (isobar model) in the sequential decay process
$J/\psi\to\gamma X, X\to\eta\eta$ are constructed using covariant tensor amplitudes described in Ref.\refcite{Zou:2002ar}.

The probability to observe the event characterized by the measurement $\xi$ is
\begin{equation}
P(\xi)=\frac{\omega(\xi)\epsilon(\xi)}{\int d\xi\omega(\xi)\epsilon(\xi)},
\end{equation}
where $\epsilon(\xi)$ is the detection efficiency and
$\omega(\xi)\equiv\frac{d\sigma}{d\Phi}$ is the differential cross
section, and $d\Phi$ is the standard element of phase space.
\begin{equation}
\frac{d\sigma}{d\Phi}=|\sum_{j}^{wave}{\Lambda_{j}A_{j}}|^{2},
\end{equation}
where $A_{j}$ is the partial wave amplitude with coupling strength
determined by a complex coefficient $\Lambda_{j}$. The normalization integral is performed
numerically by Monte Carlo techniques. The likelihood for a particular model is
\begin{equation}
\mathcal{L}=\prod\limits_{i=1}^{N_{data}}P(\xi_{i}).
\end{equation}

A series of likelihood fits are performed for parameter estimation
and model evaluation. In the log likelihood calculation, the likelihood value of
background events are given negative weights, and are removed from data since the
log likelihood value of data is the sum of signal and background.

As this involves the computation of the amplitude for every event in every iteration of
a fit, this becomes computationally very expensive for large data samples. As events
are independent and the amplitude calculation does not vary from event to event,
this task is trivially parallelizable. This and the floating point intensity predestine PWA
for implementation on graphics processing units (GPUs). GPUPWA \cite{gpupwa}\cite{GPUPWA_sourceforge} has been
developed as the working framework of BESIII, harnessing GPU parallel acceleration.
The framework now provides facilities for amplitude calculation, minimization and
plotting and is widely used for analyses at BESIII.

GPUPWA  has been developed as the working framework of BESIII, harnessing GPU parallel acceleration. GPUPWA is now developed with \emph{OpenCL} \cite{OpenCL} as described in \cite{Taipeh}.
The framework now provides facilities for amplitude calculation, minimization and plotting and is widely used for analyses at BES III. It continues to be developed and is available at \cite{GPUPWA_sourceforge}.

\begin{figure*}[htbp]
   \vskip -0.1cm
   \vspace*{8pt}
   \centering
   {\includegraphics[width=4cm,height=4cm]{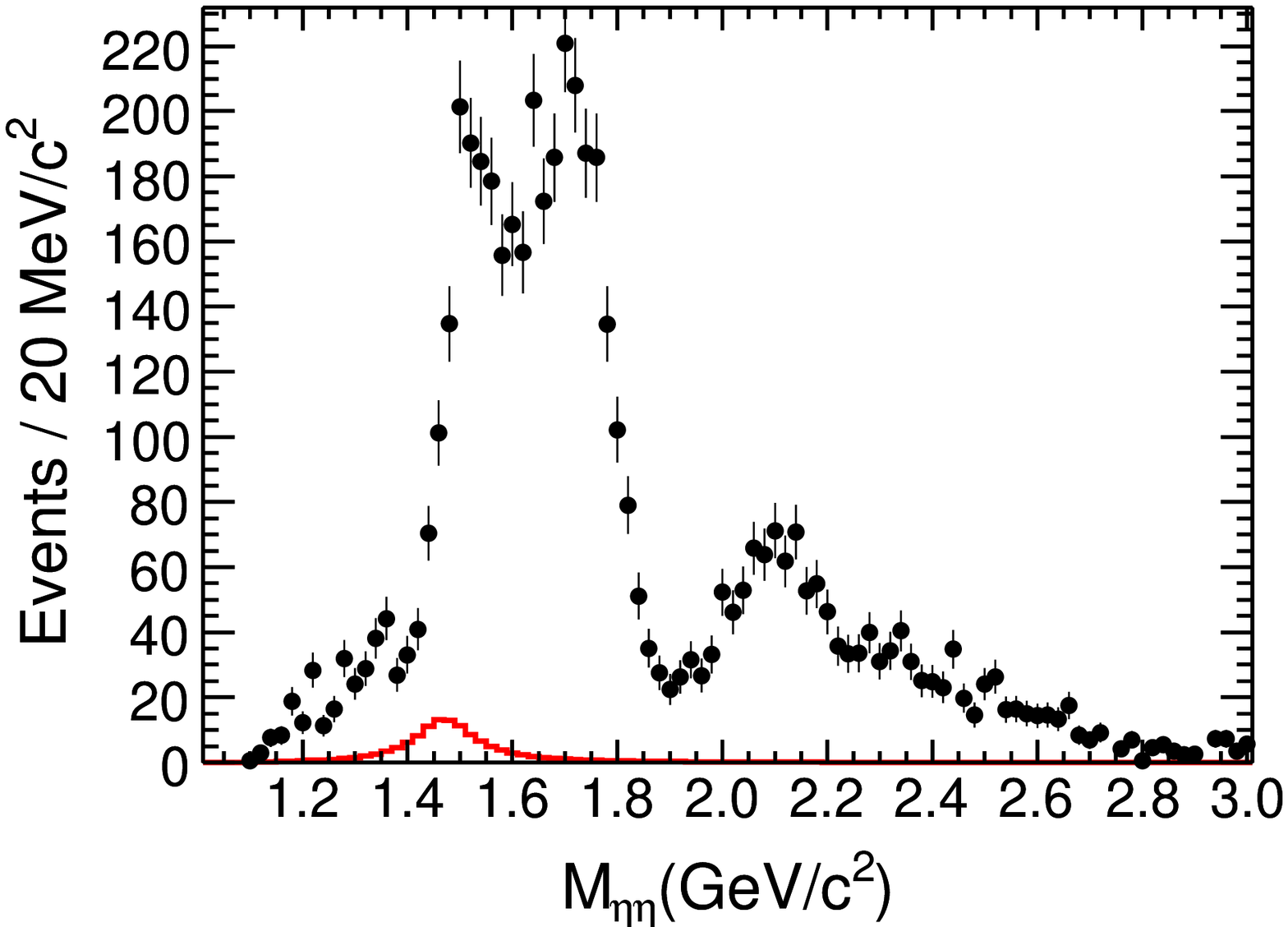}
    \put(-90,2){(a)}}
   {\includegraphics[width=4cm,height=4cm]{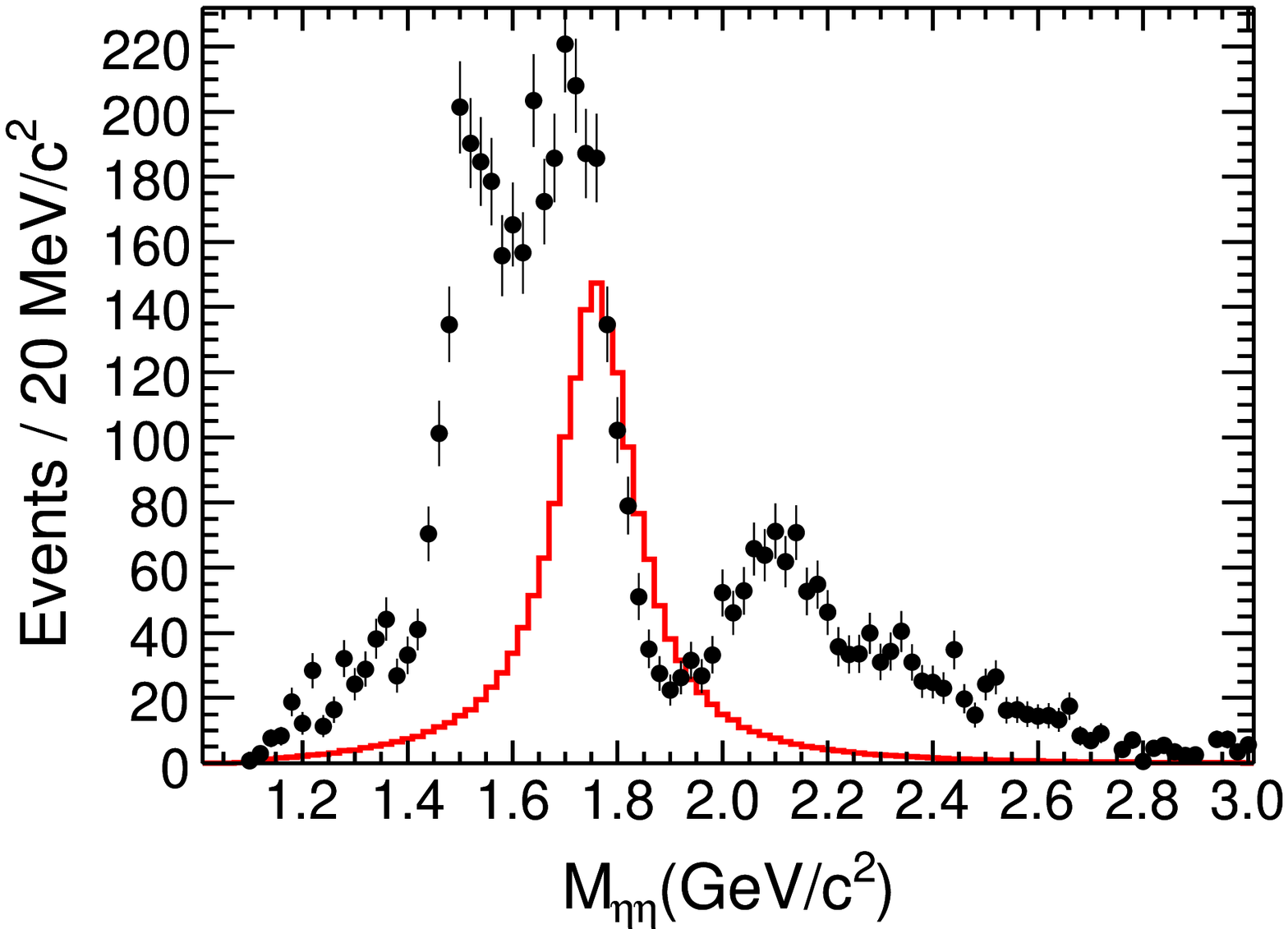}
    \put(-90,2){(b)}}
   {\includegraphics[width=4cm,height=4cm]{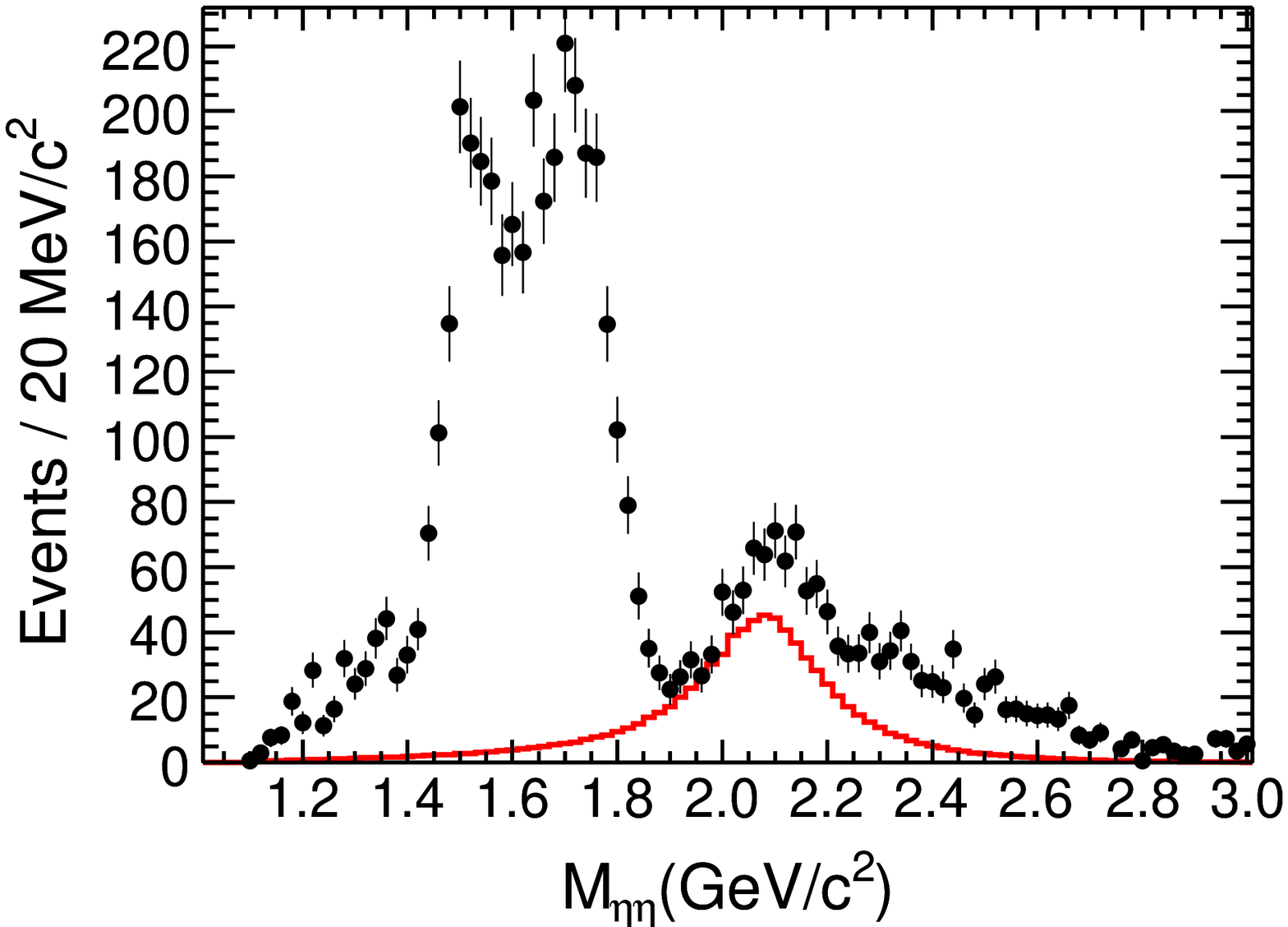}
    \put(-90,2){(c)}}
   {\includegraphics[width=4cm,height=4cm]{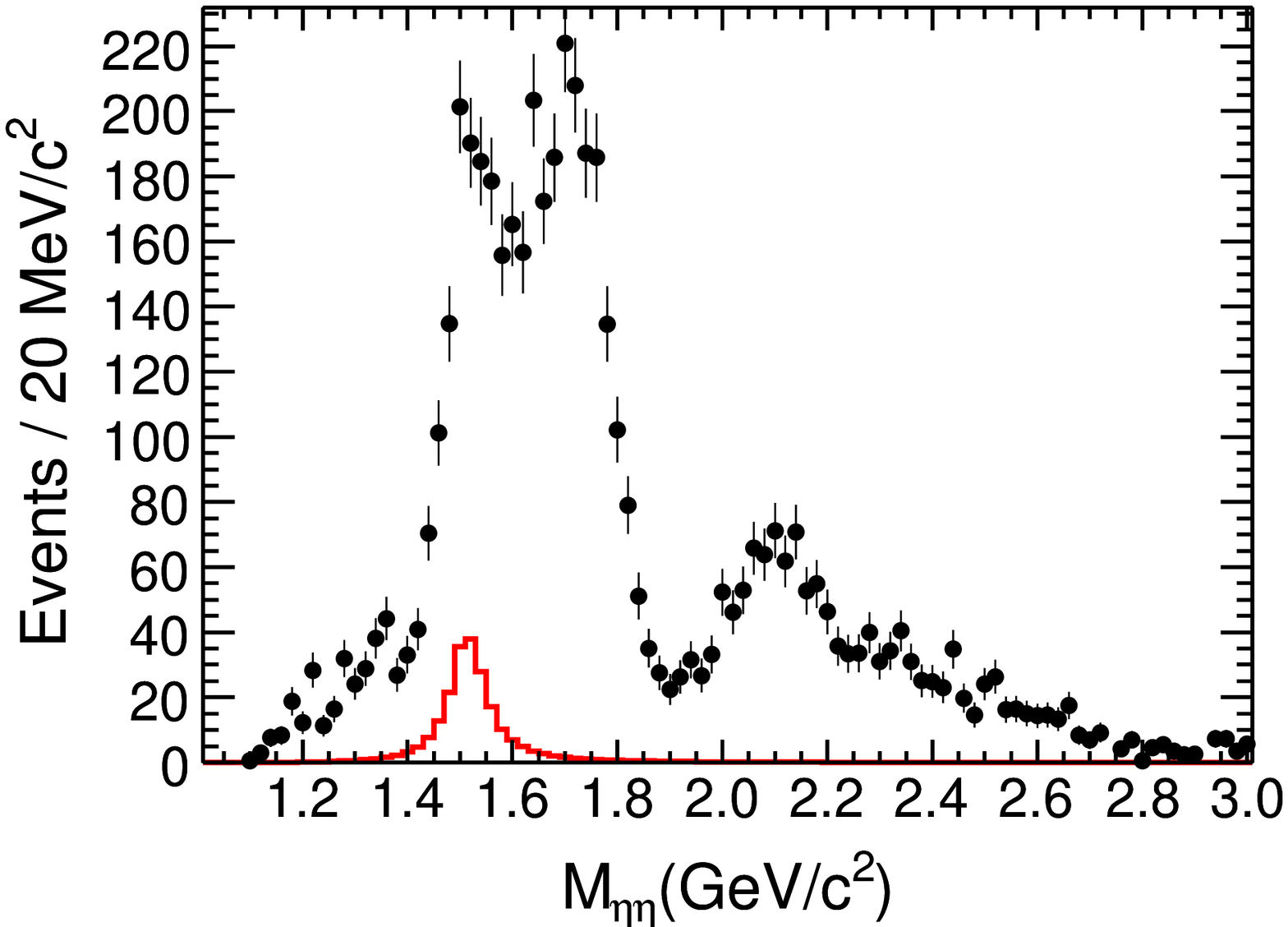}
    \put(-90,2){(d)}}
   {\includegraphics[width=4cm,height=4cm]{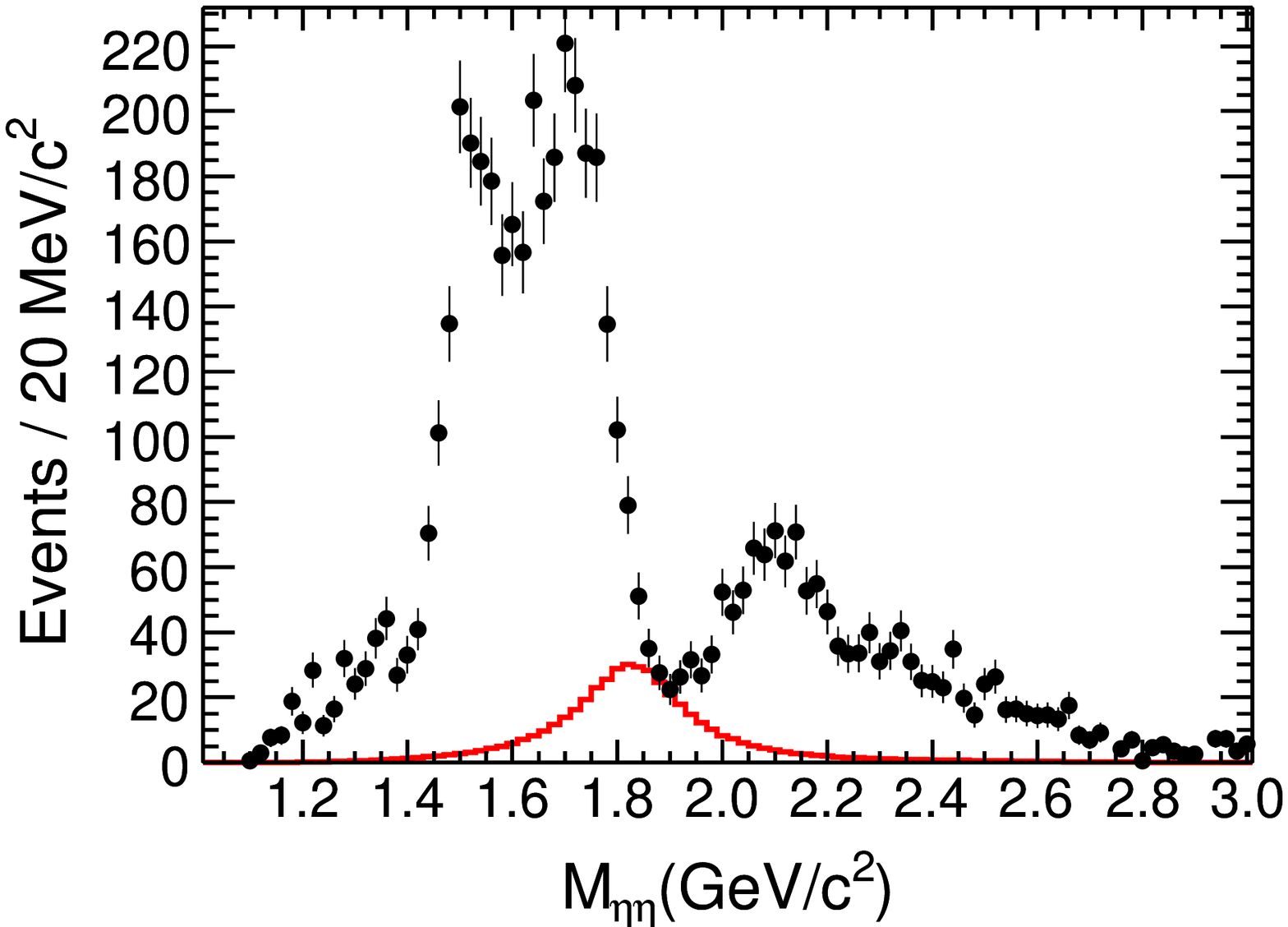}
    \put(-90,2){(e)}}
   {\includegraphics[width=4cm,height=4cm]{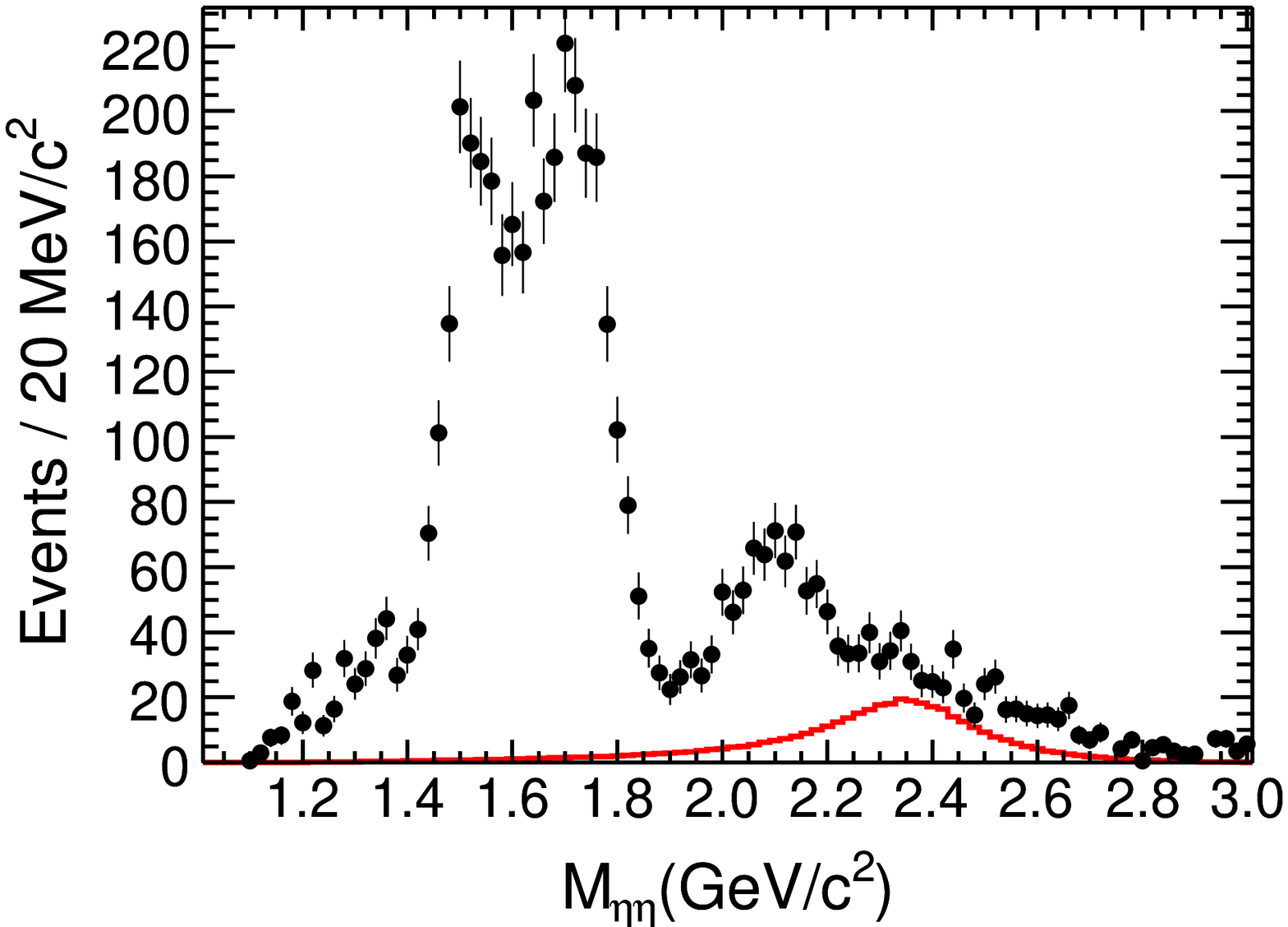}
    \put(-90,2){(f)}}
   {\includegraphics[width=4cm,height=4cm]{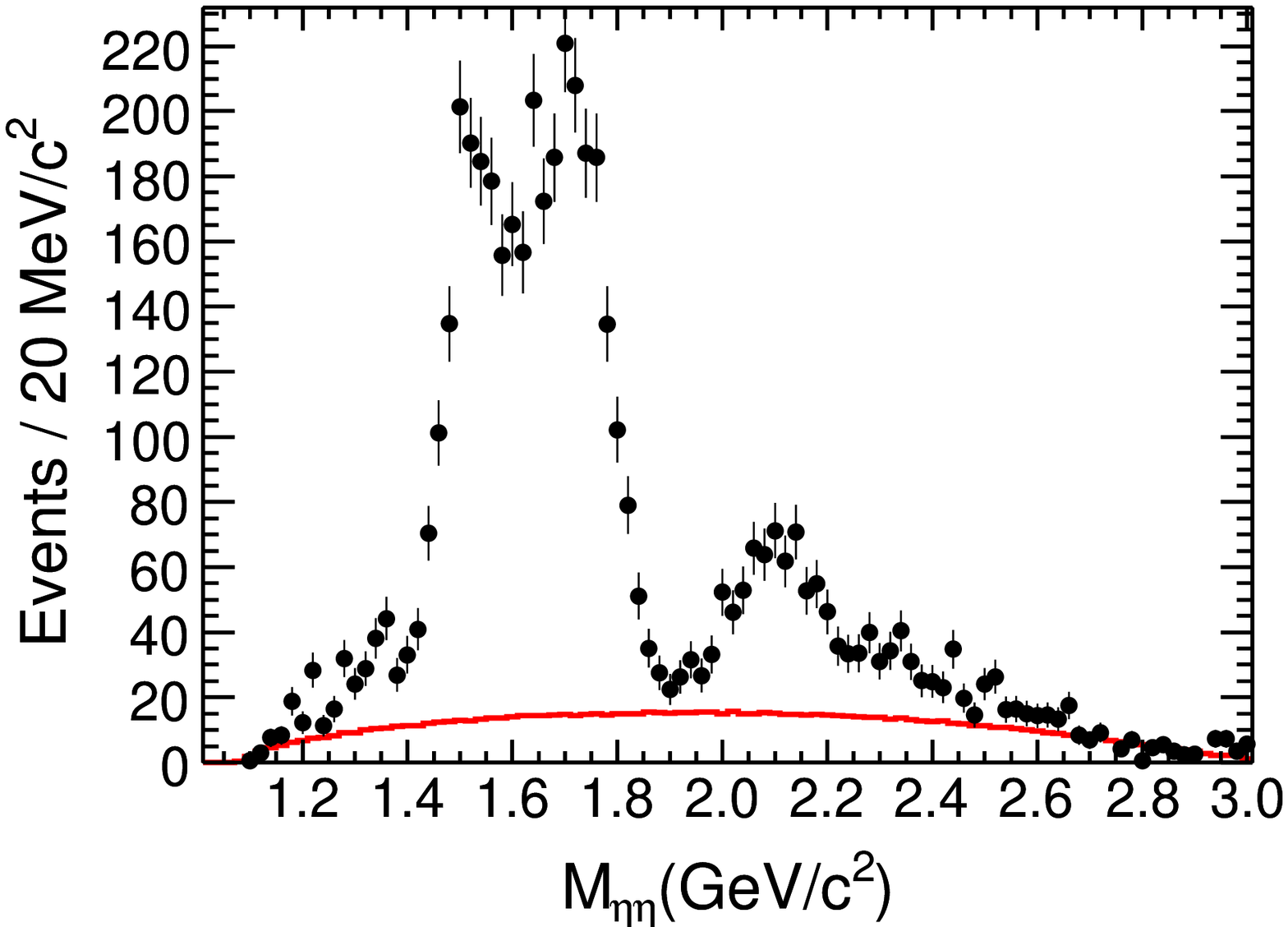}
    \put(-90,2){(g)}}
   {\includegraphics[width=4cm,height=4cm]{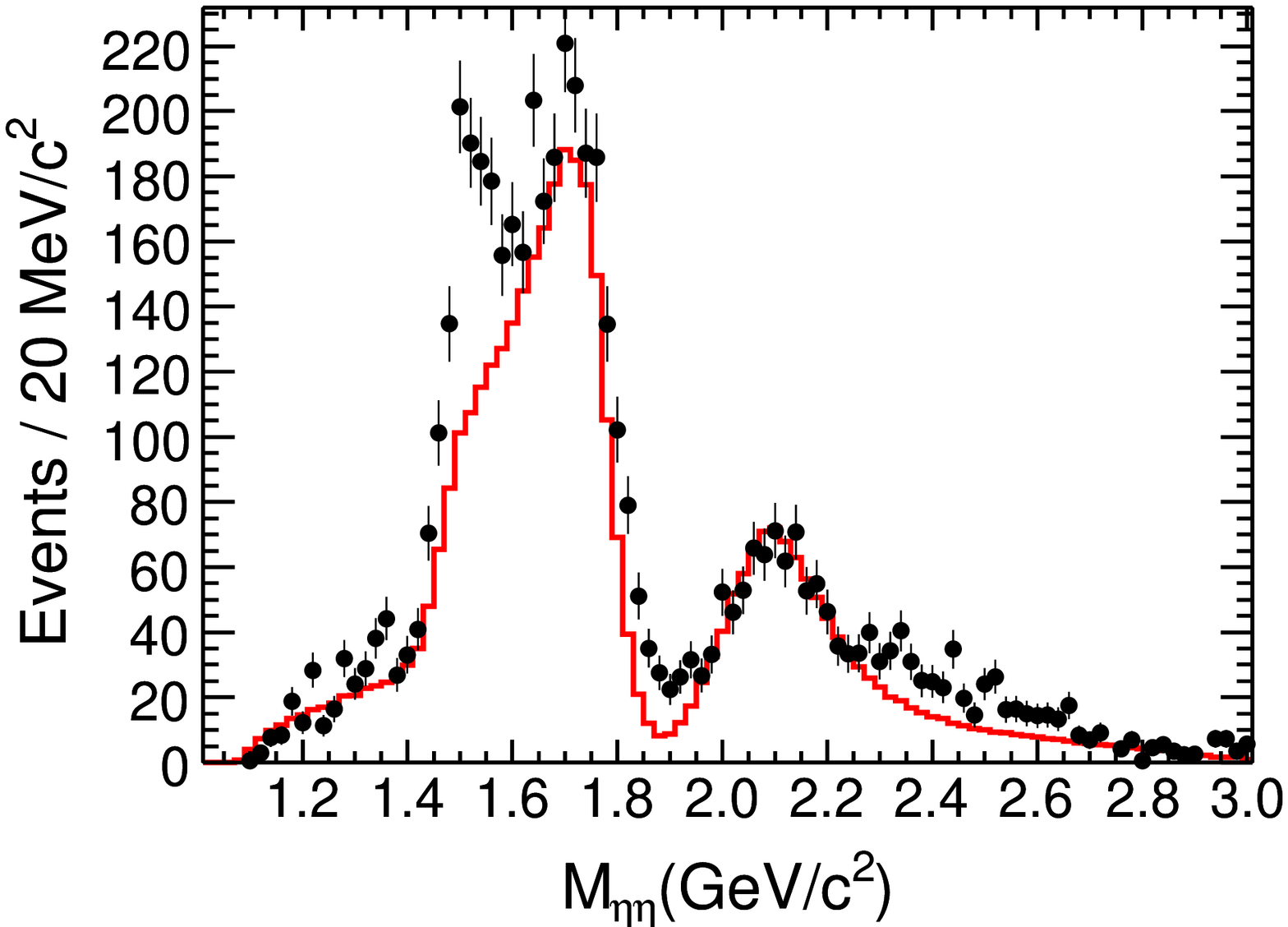}
    \put(-90,2){(h)}}
   {\includegraphics[width=4cm,height=4cm]{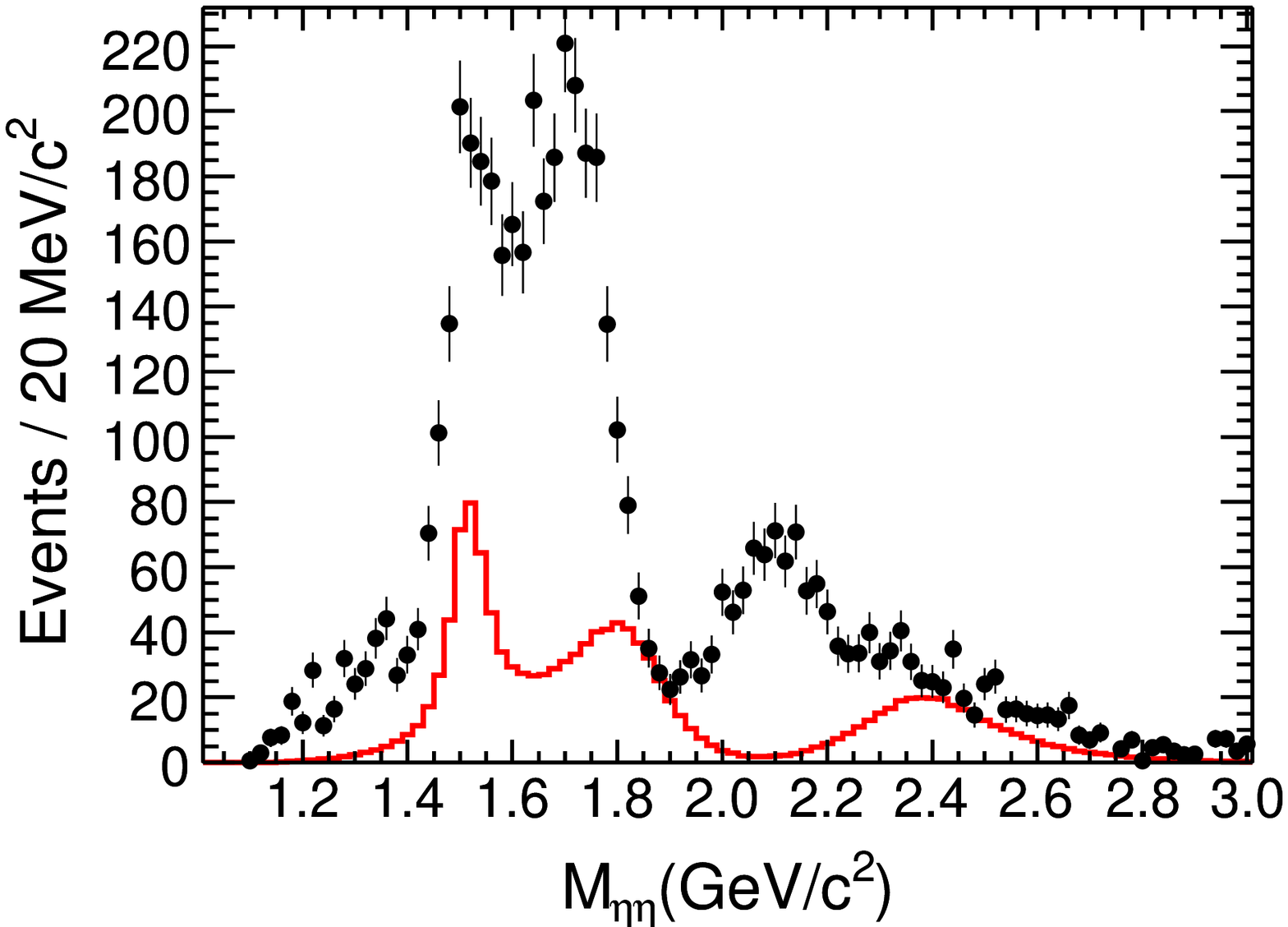}
    \put(-90,2){(i)}}

  \caption{Contribution of the components. (a) $f_{0}(1500)$, (b)
    $f_{0}(1710)$, (c) $f_{0}(2100)$, (d) $f_{2}^{'}(1525)$, (e)
    $f_{2}(1810)$, (f) $f_{2}(2340)$, (g) $0^{++}$ phase space, (h)
    total $0^{++}$ component, and (i) total $2^{++}$ component.  The dots
    with error bars are data with background subtracted, and the solid
    histograms are the projection of the PWA results\protect\cite{Ablikim:2013hq}.}
   \vskip 10pt
   \label{fig:component}
\end{figure*}

\section{PWA results of $J/\psi\to\gamma\eta\eta$}

For a $J/\psi$ radiative decay to two pseudoscalar mesons, it offers a very clean
laboratory to search for scalar and tensor glueballs because only intermediate states
with $J^{PC} = even^{++}$ are possible. An early study of $J/\psi\to\gamma\eta\eta$ was made by the Crystal
Ball Collaboration with the first observation of $f_{0}(1710)$, but the study suffered from
low statistics. The results of partial wave analysis (PWA) on $J/\psi\to\gamma\eta\eta$ (Fig.~\ref{fig:component}) are presented based on a sample of $2.25\times10^8$ $J/\psi$ events collected with BESIII \cite{Ablikim:2013hq}.

\section{Summary and outlook}

A full partial wave analysis was
performed to disentangle the structures
present in $J/\psi\to\gamma\eta\eta$ decays. The scalar
contributions are mainly from $f_{0}(1500)$,
$f_{0}(1710)$ and $f_{0}(2100)$, while no evident
contributions from $f_{0}(1370)$ and $f_{0}(1790)$
are seen. Recently, the production rate
of the pure gauge scalar glueball in J/£r
radiative decays predicted by the
lattice QCD \cite{Gui:2012gx} was found to be
compatible with the production rate of
$J/\psi$ radiative decays to $f_{0}(1710)$; this
suggests that $f_{0}(1710)$ has a larger
overlap with the glueball compared to
other glueball candidates (eg. $f_{0}(1500)$).
In this analysis, the production rate of $f_{0}(1710)$ and $f_{0}(2100)$ are both about one
order of magnitude larger than that of
the $f_{0}(1500)$, which are both consistent
with, at least not contrary to, lattice
QCD predictions \cite{Gui:2012gx}.

Now five years from our first collisions,
BESIII has established a broad and
successful program in charm physics.
Recently, in 2012, even larger samples
have been accumulated at the $J/\psi$ and
$\psi(3686)$; total samples are now about 1.2
billion and 0.35 billion decays,
respectively. Furthermore, our 2013
dataset includes more data near 4260
MeV, and also a large sample at the
$Y(4360)$. With the excellent
performance of the accelerator and
detector, more interesting results are
expected.

%\begin{thebibliography}{000} %for 3 digits
%\begin{thebibliography}{00}  %for 2 digits

\end{document}